# Voltage-Programmable Photon Statistics Using a High-Extinction Thin-film Lithium Niobate Modulator

*Julian Rasmus Bankwitz[1,\*], Ravi Pradip[1], Julius Römer[1], Frank Brückerhoff-Plückelmann[1], Falk Ebert[1], Lennart Meyer[1], Liam McRae[1], Jan Brandes[1], Akhil Varri[1,2], Wladick Hartmann[3], Wolfram H.P. Pernice[1,2,\*], and Xinyu Ma[1,\*]*

[1]Kirchhoff-Institute for Physics, Heidelberg University, Im Neuenheimer Feld 227, 69120 Heidelberg, Germany

[2]Physics Institute, University of Münster, Heisenbergstraße 11, 48149 Münster, Germany

[3]Pixel Photonics GmbH, Heisenbergstraße 11, 48149 Münster, Germany

[\*]   julian_rasmus.bankwitz@kip.uni-heidelberg.de;   wolfram.pernice@kip.uni-heidelberg.de;   xinyu.ma@kip.uni-heidelberg.de

## ABSTRACT

**Controlling the statistical properties of light, namely the fluctuations in photon arrival, entropy and number, is essential for both classical and quantum photonics. While integrated systems provide tunable control over amplitude, phase, and wavelength, real-time modulation of photon statistics has remained a long-standing challenge. Herein, we introduce the concept and experimental realization of a photon statistics transducer: a high-extinction, broadband electro-optic device capable of deterministically shaping photon-number distributions at nanosecond timescales. Our approach employs a cascaded thin-film lithium niobate (TFLN) Mach–Zehnder amplitude modulator delivering >50 dB extinction, enabling precise suppression and release of coherent seed light from an integrated InP laser. By exploiting the interplay between seed suppression and erbium-doped fiber amplifier dynamics, we demonstrate smooth, voltage-controlled switching between Poissonian and super-Poissonian photon statistics, with second-order coherence $g^{(2)}(0)$ tunable from 1.0 to 1.7. Complementary measurements with superconducting nanowire single-photon detectors further show photon-flux control down to sub-photon levels, highlighting the potential for future operation with non-classical sources. The photon statistics transducer thus establishes statistical modulation as a new functional primitive in integrated photonics. Applications range from entropy generation and secure communication to neuromorphic and hybrid quantum-classical processing, where controlled randomness and entropy are essential resources. By enabling programmable transitions between statistical regimes using only electronic**

**drive signals, our work lays the foundation for adaptive, entropy-aware photonic systems that bridge classical and quantum domains.**

## Introduction

The temporal statistics of photon arrival events, including their correlations, fluctuations, and multimode structure, play a central role in classical and quantum optical systems because they determine noise propagation, intensity correlations $g^{(2)}(0)$, and the photon-number distributions that define optical entropy[1–7]. Photon statistics not only reflect the quantum or thermal nature of a light source, but also govern how information, noise, and entropy propagate through optical systems[3,8–12]. Poissonian light, such as that produced by ideal lasers, exhibits uncorrelated photon arrival and low noise, while super-light, typical of thermal or amplified spontaneous emission (ASE) sources, displays photon bunching and enhanced fluctuations[9,13,14]. These distinct regimes are not mere descriptors; they are functional resources for modern photonic applications, including neuromorphic computing[2,9,15–19], physical random number generation[20,21], secure optical communication[22], and hybrid quantum-classical information processing[23,24] (Figure 1a).

Despite this emerging relevance, the ability to dynamically modulate photon statistics in real time has remained elusive. Conventional methods rely on static source engineering or passive post-processing, both of which are slow, lossy, and incompatible with the requirements of integrated photonics[9,10,14,20]. Today's reconfigurable optical systems can readily manipulate phase[25], amplitude[26–28], wavelength[29,30], and spatial mode[31], but the statistical structure of light remains fixed during operation. This has created a fundamental bottleneck: many photonic systems now process information in real time and rely on stochastic optical signals whose properties are fixed by the underlying noise source[32,33]. Several of these architectures operate within narrow temporal windows, and in some cases these windows reach the tens-of-picoseconds regime[9]. In such scenarios, the ability to tune photon-number fluctuations on comparable timescales provides a useful control parameter that fixed-entropy sources cannot offer[2,14,20].

Beyond the limiting cases of coherent ($g^{(2)}(0) = 1$) and thermal ($g^{(2)}(0) = 2$) light, the intermediate regime $1 < g^{(2)}(0) < 2$ carries distinct physical meaning: it reflects a controlled mixture of stimulated and spontaneous emission and therefore a tunable level of optical entropy. Continuous control over this regime provides functionality that binary switching alone cannot offer[2,9,20]. Adjustable stochasticity plays a key role in photonic reservoir computing and neuromorphic processors, where noise strength influences nonlinear activation dynamics and memory capacity[9,15]. Likewise, controlled intermediate fluctuations support statistical masking in secure links, photonic Monte-Carlo sampling, and tunable stochastic behavior in hybrid quantum–classical systems[2,9,22,34]. These applications require graded noise levels rather than only the limiting coherent and thermal states.

In this work, we introduce a photon statistics transducer: a voltage-programmable, integrated electro-optic device that deterministically shapes photon-number distributions in real time. In this manuscript, we use the term photon statistics transducer to denote a device in which an electrical control signal deterministically sets the photon-number distribution of an optical field, while the underlying optical source remains unchanged. The device therefore performs a mapping from voltage to statistical regime, enabling programmable transitions between Poissonian and super-Poissonian photon statistics. The transducer is enabled by a dual-stage TFLN Mach-Zehnder amplitude modulator providing >50 dB extinction, which permits high-dynamic-range gating of coherent light on nanosecond timescales. With this capability, we realize electronic switching between Poissonian and super-Poissonian regimes, continuously tuning the second-order coherence from $g^{(2)}(0) = 1.0$ to 1.7, and we demonstrate precise control of photon flux down to sub-photon levels. All constituent elements Indium Phosphide (InP) laser, TFLN amplitude modulator and Erbium amplifier (see Figure 1b) are compatible with wafer-scale photonic integration, positioning the approach as a scalable building block for programmable optical systems. Figure 1c and d show optical micrographs of the used foundry level TFLN modulator an InP Laser.

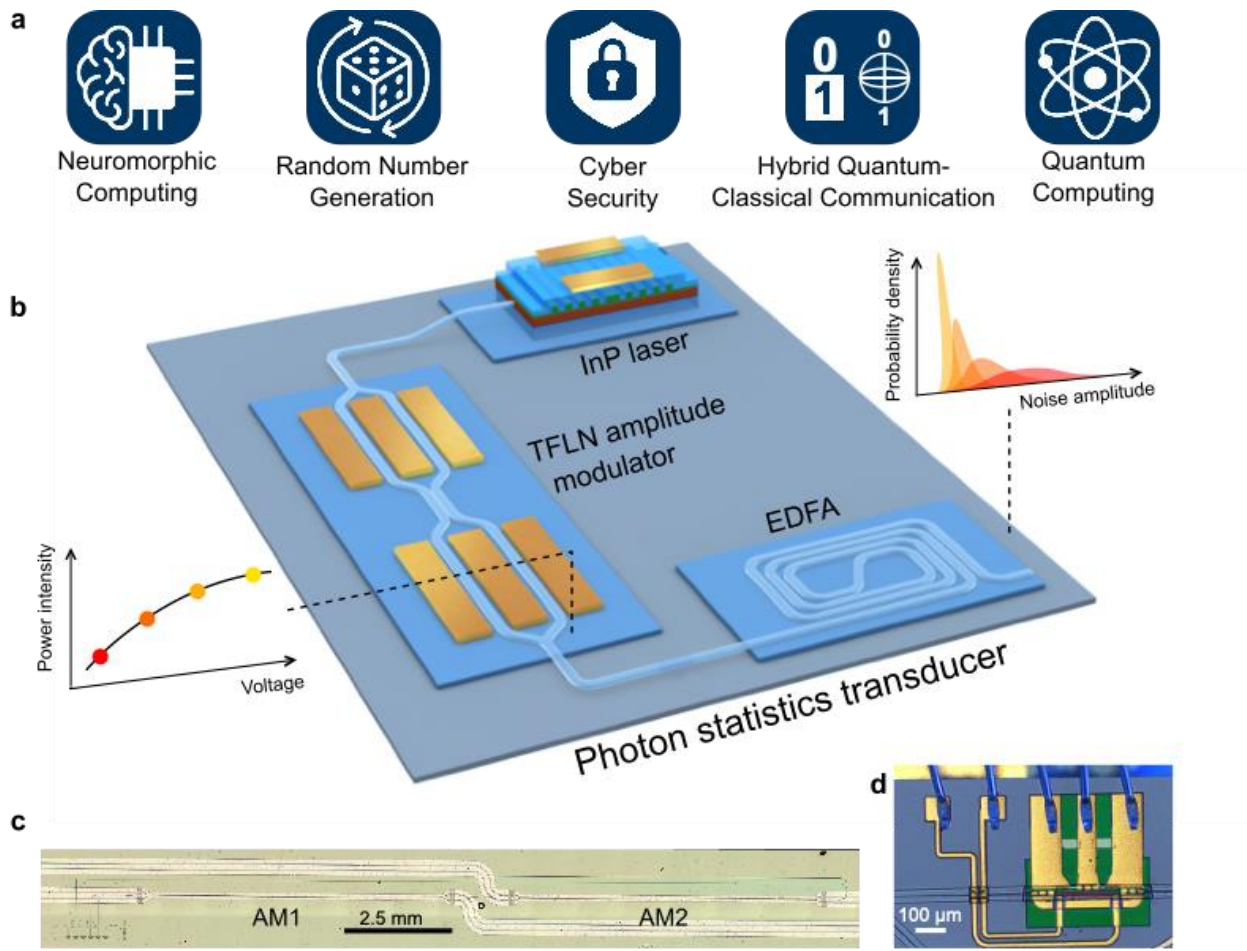

**Figure 1. Concept of a Photon Statistics Transducer** a) Applications of the photon statistics transducer b) Schematic of the photon-statistics transducer. A continuous wave InP (indium phosphide) laser is intensity modulated by a dual stage thin-film lithium niobate (TFLN) amplitude modulator whose transmission is set by the applied voltage. The modulated light is then amplified by an erbium-doped fiber amplifier (EDFA), yielding output photon-number statistics determined by the voltage-dependent transmission. c) Optical micrograph of an integrated dual stage TFLN amplitude modulator (AM) with input/output grating couplers on the lower left d) Optical micrograph of an integrated InP Laser, including wavelength tuning heater.

Programmable statistical modulation opens a complementary control axis to phase, amplitude, and wavelength. In hybrid quantum and classical links, it enables statistical multiplexing with dynamic, on demand transitions between entropy rich and deterministic segments within a single channel[4,5,32,35]. In neuromorphic and reservoir computing, it provides a fast, tunable optical noise source for stochastic processing and learning[2,9]. For security and entropy generation, it functions as an electrically driven randomness engine with nanosecond agility[34,36]. By elevating photon-statistics control to a first-class knob, the transducer lays the groundwork for adaptive, entropy-aware photonic architectures that bridge classical and quantum domains while remaining fully integration-ready.

## Results

**Device design and working principle**

The temporal structure of photon arrival statistics, whether Poissonian, bunched, or antibunched, can be shaped by precisely modulating the transmission envelope of a light field. High-extinction amplitude modulation[37,38] is key to ensuring that statistical transformations are governed solely by the modulation waveform, free from background leakage that would otherwise mask fine correlations. To achieve high-speed, high-fidelity switching between distinct photon-number statistics, we designed a broadband, electro-optic amplitude modulator architecture with exceptional extinction performance and robustness to fabrication imperfections. While TFLN modulators are known for their ultra-wide electrical bandwidths, often exceeding 100 GHz[39], conventional integrated designs typically exhibit extinction ratios limited to 30 dB[40]. This constraint arises primarily from fabrication-induced deviations in directional coupler splitting ratios, which prevent ideal interference conditions and thus limit the achievable on/off contrast in Mach-Zehnder interferometer (MZI) structures[41].

To overcome this limitation, we developed a cascaded architecture composed of two serial TFLN amplitude modulators, connected via a 2×2 directional coupler, as shown in Figure 2a. Each modulator consists of a standard MZI with phase modulators in both arms, driven in push-pull configuration. The central 2×2 coupler interferometrically links the two AMs, forming a coherently coupled system in which field amplitudes are redistributed between the bar and cross output ports. This structure provides multiple degrees of freedom that enable passive error compensation: by effectively correcting imperfections across three couplers $S_1, S_2$ and $S_3$, the system can suppress residual transmission to well beyond the 50 dB threshold.

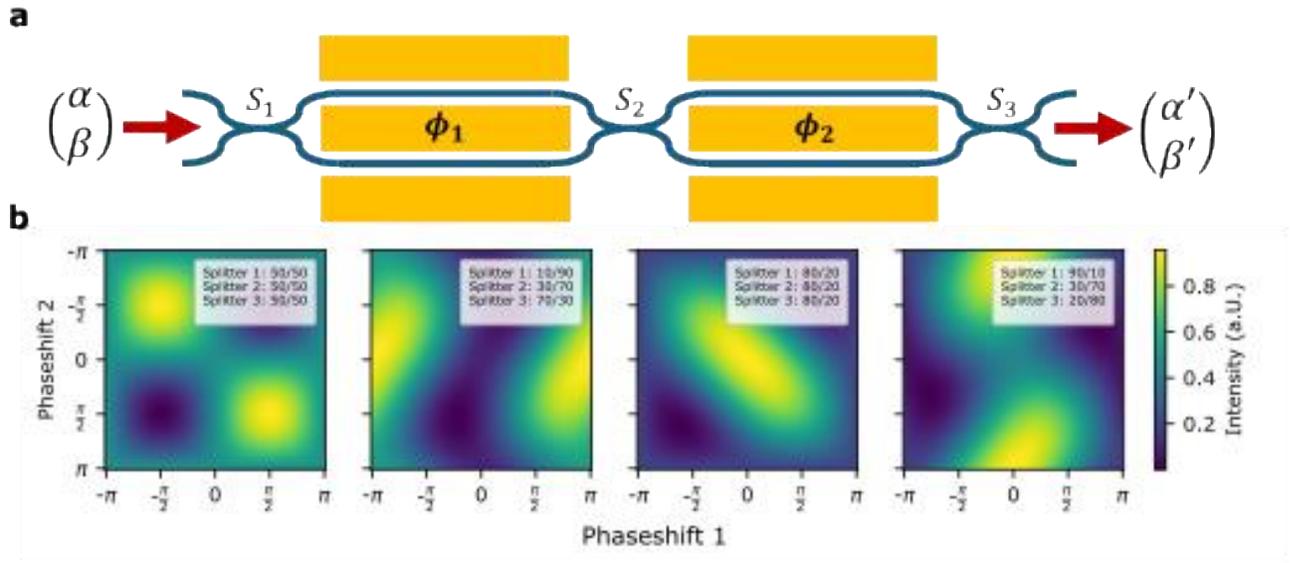

**Figure 2. Concept of fabrication resilient amplitude modulator** a) Schematic of the TFLN amplitude modulator that shows two EOMs, coupled by a 2x2 splitter. Splitting ratios $S_j$ and modulator phase shifts $\phi_j$ are indicated. The optical field amplitudes at in-/output of the device are $\alpha\binom{}{}$ and $\beta\binom{}{}$ b) Optical output field amplitude $\alpha'$ depending on the phase shifts applied to the EOMs plotted for different splitting ratios $S_{1,2,3}$. Each of the combination exceeds extinction ratios of 50 dB, indicating fabrication resilience of the amplitude-modulators performance.

The optical behavior of this system can be described analytically using the transfer matrix formalism[41]. Each directional coupler $S_j$ is represented by a unitary matrix with transmission and reflection coefficients $t_j$ and $r_j$, while each electro-optic modulator $\phi_j$ introduces a phase shift $\varphi_j$ to each arm:

$$S_j = \begin{bmatrix} \sqrt{t_j} & i\sqrt{r_j} \\ i\sqrt{r_j} & \sqrt{t_j} \end{bmatrix} \quad (1)$$

$$\phi_j = \begin{bmatrix} e^{i\varphi/2} & 0 \\ 0 & e^{-i\varphi/2} \end{bmatrix} \qquad (2)$$

the total transfer function of the system is then given by:

$$\begin{bmatrix} \alpha' \\ \beta' \end{bmatrix} = S_3 \phi_2 S_2 \phi_1 S_1 \begin{bmatrix} \alpha \\ \beta \end{bmatrix} \qquad (3)$$

where $\alpha, \beta$ are the input field amplitudes, and $\alpha', \beta'$ are the outputs. By sweeping the internal phase shifts $\phi_1$ and $\phi_2$, one can achieve destructive interference at the output ports even in the presence of nonideal coupler parameters. Simulations of the transfer function over a wide range of directional-coupler splitting ratios confirm that extreme extinction is theoretically achievable with appropriate phase settings, irrespective of fabrication offsets (Figure 2b) (For detailed mathematical proof see Supplementary Section 1). This tolerance is especially relevant in TFLN, where high-efficiency modulators rely on partially etched waveguides: unavoidable etch-depth variations perturb the effective indices and coupling gaps, leading to unpredictable directional-coupler ratios. The architecture thus provides inherent resilience to coupler asymmetry, a long-standing limitation in integrated TFLN photonic circuits. Figure 2b shows the simulated output amplitude $\alpha'$, as a function of applied phase bias, for various non-ideal splitter configurations. As shown, the interferometric structure maintains high contrast even with asymmetric coupler ratios. Experimentally, we demonstrate an extinction ratio of 51 dB, as shown in Figure 3, by adjusting the RF drive voltages applied independently to each modulator. The upper two plots of Figure 3a show the normalized, measured transmissions $\alpha'$ and $\beta'$, whereas the lower two are the result of a plane fit of Equation 3 on the measured results. The fit reveals the transfer splitting matrices (Equation 1) of the transfer function Equation 3 of the device to be

$$S_1 = \begin{bmatrix} \sqrt{0.27} & i\sqrt{0.73} \\ i\sqrt{0.73} & \sqrt{0.27} \end{bmatrix}, S_2 = \begin{bmatrix} \sqrt{0.71} & i\sqrt{0.29} \\ i\sqrt{0.29} & \sqrt{0.71} \end{bmatrix} \text{ and } S_3 = \begin{bmatrix} \sqrt{0.8} & i\sqrt{0.2} \\ i\sqrt{0.2} & \sqrt{0.8} \end{bmatrix} \qquad (4)$$

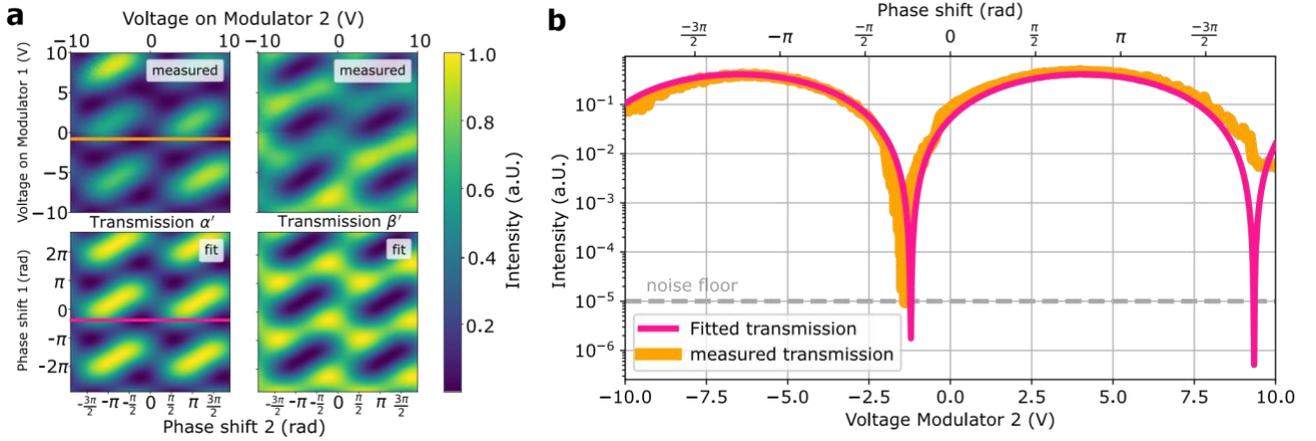

**Figure 3. Voltage depended transmission map** a) Transmission $\alpha'$ and $\beta'$ at output of TFLN amplitude modulator. The top row shows measured results in dependance of the voltage applied to EOM 1 and EOM 2. The maximum extinction ratio measured is 51 dB. The bottom row shows a plane fit of the measurement data using Equation 4. Fitted parameters from the transmission measurements of the transducer are displayed in Equation 4, Equation 5, and Equation 6 b) Line extraction of measured and fitted transmission along the colored lines indicated in a). The line was chosen to pass through the minimum transmission point, which equals 51 dB extinction, equal to the noise floor of our measurement setup. The curve of the fitted transmission, indicates potential for around 60 dB of our fabrication resilient amplitude modulator.

The impact of fabrication tolerances in a wafer-scale foundry process is evident: although all splitters were nominally designed for identical coupling ratios, the measured devices exhibit only similar values, with a mean 75/25% splitting ratio and a standard deviation of 4 %. Achieving an extinction ratio exceeding 50 dB despite these unbalanced splitters highlights the robustness of the fabrication-tolerant design. The voltage depended phase shift (Equation 2) induced by the modulator from Equation 3 is fitted to be

$$\phi_1 = 0.88 \frac{\mathrm{V}}{\mathrm{rad}} - 0.96 \text{ rad} \tag{5}$$

$$\phi_2 = 0.60 \frac{\mathrm{V}}{\mathrm{rad}} + 1.30 \text{ rad} \tag{6}$$

where the phase shifts get an offset added, to take MZI imbalances into account. In Figure 3b, the transmission is plotted along the colored line in Figure 3a. The line is chosen to pass the maximum extinction point that is reaching 51 dB, equivalent to the noise floor of our measurement setup. However, this performance level significantly exceeds that of standard monolithic modulators and is critical for the high-dynamic-range operation of the photon statistics transducer.

**Photon statistics modulation**

This high-extinction, electrically broadband performance allows us to dynamically shape the photon statistics of light using fast attenuation. As shown in Figure 2, the modulated output is routed into a low-noise erbium-doped fiber amplifier (EDFA), where the statistical regime of the light depends on the instantaneous power (Figure 1). When the modulator is fully open, the EDFA output is dominated by stimulated emission, producing narrowband coherent light with Poissonian statistics. When the seed is suppressed by driving the modulator into its extinction point, the EDFA emits broadband amplified spontaneous emission (ASE), which exhibits thermal photon statistics and super-Poissonian noise. Reaching the extinction point requires an amplitude modulator on-off suppression sufficient to push the residual seed at the EDFA input below the ASE level within the detection bandwidth; in our setup this corresponds to >50 dB optical suppression. Intermediate driving voltages allow smooth tuning between these two regimes.

This seed-on/seed-off behavior produces a clear statistical transition in the output light, even though the laser wavelength and all downstream components remain unchanged. To analyze the resulting intensity fluctuations, we measured the root mean square (RMS) voltage from a high-speed photodiode placed at the EDFA output. Figure 4a shows the measured RMS noise amplitude as a function of modulator drive voltage, revealing a continuous transition in optical noise level that mirrors the change in photon statistics. The highest noise is observed when the EOM transmission is fully suppressed, corresponding to pure ASE output. As the modulator opens and allows increasing amounts of coherent light through, the noise amplitude decreases, reaching a minimum in the stimulated-emission-dominated regime.

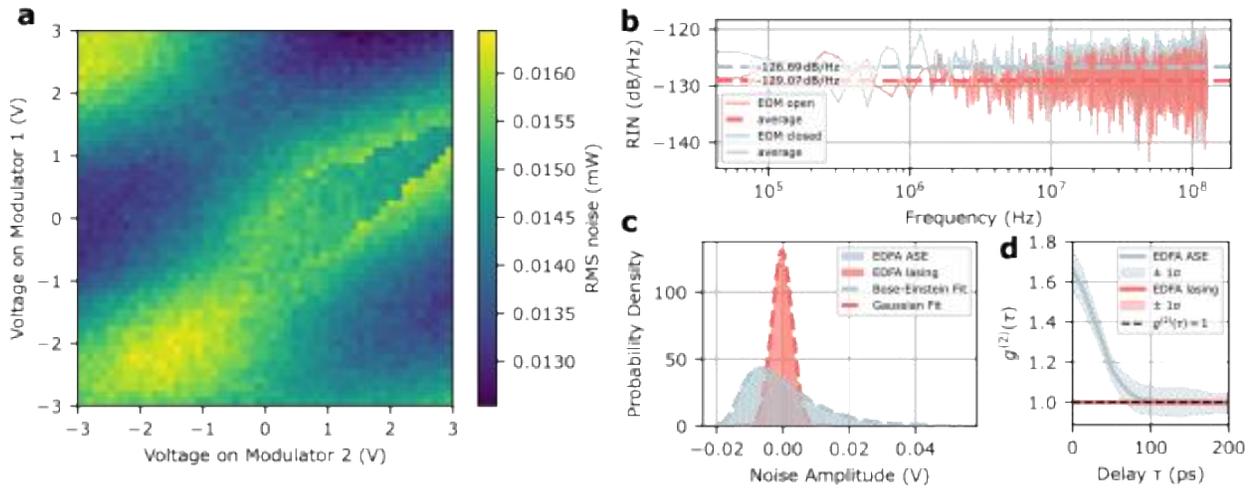

**Figure 4. Photon statistics at EDFA output** a) RMS noise of EDFA output in dependence of voltage applied to the two EOMs of the amplitude modulator. The RMS noise is low, when the amplitude modulator output is high and vice versa. This behavior arises from the EDFA being in noisy, spontaneous emission, when the EOM does not transmit light, while it is in its less noisy, stimulated emission state, when the light amplitude is high. b) Relative intensity noise of the minimum and maximum values of the RMS noise. The dashed lines indicate the mean of the RIN. The RIN of the EOM open (EDFA lasing) is 2.38 dB/Hz lower than the EOM open state (EDFA spontaneous emission). c) Noise amplitude of the two different output modes from b). When the EDFA is lasing, the light follows a Gaussian distribution, whereas it follows a convolution of an M-fold Bose-Einstein distribution and a Gaussian distribution, when the EDFA is in ASE mode. d) Autocorrelation measurement of the two states. The lasing state shows $g^{(2)}(0) = 1$ indicating coherence, whereas the ASE mode results in $g^{(2)}(0) = 1.661 \pm 0.098$ indicating photon bunching, typical for thermal light sources like an EDFA ASE.

To confirm that the observed noise modulation arises from a statistical change in the light field and not from trivial variations in output power, we performed relative intensity noise (RIN) measurements in both the open and closed modulator states. These measurements were taken directly at the photodiode output, downstream of the EDFA. As shown in Figure 4b, the difference in RIN spectra between the two states amounts to $\Delta_{RIN}$ = 2.38 dB/Hz, demonstrating a clear increase in normalized intensity noise when the light is dominated by spontaneous emission. This observation is further supported by histograms of the detector voltage, shown in Figure 4c, which reveal a transition from a narrow Gaussian distribution in the lasing case to a broader, asymmetric profile in the noise-dominated regime. The latter is consistent with a convolution of M-fold Bose-Einstein statistics from the thermal ASE source and the inherent Gaussian shot noise of the detector. It is to be mentioned, that the shot noise at the detector can be considered negligible against the noise spectrum presented in Figure 4 since the relative intensity shot noise at 0.2 mA corresponds to -147 dB/Hz. The fitted parameters of the convolution suggest a two-fold Bose-Einstein distribution (see Supplementary Section 2), which is the expected result when two statistically independent thermal modes contribute to the detected signal. This is consistent with typical amplified spontaneous emission (ASE) from an EDFA, where orthogonal polarization modes or uncorrelated temporal modes dominate the emission. Since the detector integrates over both, the observed noise statistics correspond to an *M* = 2 multimode thermal distribution.

Since the photodetector used in this experiment is AC-coupled, the resulting voltage distribution is centered around zero. As a result, negative values appear in the noise histograms even though the optical intensity is strictly positive. This zero-centering also affects the measurement of higher-order correlations. Nevertheless, we extracted the second-order intensity autocorrelation $g^{(2)}(0)$ (see Figure 4d) from the photodiode voltage traces, sampled at 80 GSa/s using a 38 GHz detector. In the coherent regime (modulator open), we measured $g^{(2)}(0) = 1.001 \pm 0.001$, consistent with uncorrelated Poissonian fluctuations. In the ASE-dominated regime (modulator closed), the measured autocorrelation rises to $g^{(2)}(0) = 1.661 \pm 0.098$, indicating clear photon bunching characteristic of thermal light.

The deviation from the ideal value $g^{(2)}(0) = 2$, expected for chaotic light, can be attributed to the finite measurement bandwidth. The measured autocorrelation is set by the convolution of the ASE optical bandwidth and the photodiode's 38GHz electrical bandwidth. For these $g^{(2)}(\tau)$ measurements we intentionally filtered the ASE to 10 GHz bandwidth, corresponding to ≈ 80 pm at 1550 nm, so that the optical bandwidth is narrower than the detector bandwidth, thereby preserving temporal contrast in the intensity fluctuations. Nonetheless, the observed increase in $g^{(2)}(0)$, combined with the RIN and RMS data, robustly supports the conclusion that the photon statistics transducer is capable of real-time, voltage-controlled noise-domain modulation of light. This represents an important functional primitive for entropy management, secure communication, and hybrid analog-digital optical signal processing.

**Single-photon number control**

Although the experiments in this work focus on controlling classical photon statistics, we also tested the transducer in the single-photon regime to assess its potential for quantum applications. In our system, light from a coherent InP laser is attenuated to the single-photon level using the amplitude modulator with the EDFA removed. The output is then directed to a superconducting nanowire single-photon detector (SNSPD) model DENA Desktop Format by Pixel Photonics, allowing us to monitor photon statistics within user-defined time windows of $\Delta t$ = 100 μs. By varying the applied voltage to the modulator, we tune the mean photon number $\langle n \rangle$ per 100 μs interval from below 0.02 photons (limited by the systems dark count rate) to over 15 photons (limited by SNSPD latching behavior). Figure 5a, shows a logarithmic plot of the measured average photon number depending on the 2D-voltage plane.

Figure 5b presents representative histograms with Poisson fits across modulation voltages. The data is consistent with Poisson statistics over the full range, and for large means the Poisson distribution is indistinguishable from its Gaussian approximation. At low optical powers, we observe Poissonian statistics, with count histograms matching the theoretical distribution expected from an ideal laser source. The data demonstrates smooth, voltage-controlled evolution of the distribution shape, from sparse, discrete photon events to dense, normally distributed fluctuations. This behavior directly supports physical random number generation, where the randomness arises from intrinsic quantum fluctuations in photon arrival times. The transducer thus provides a voltage-tunable entropy source for quantum communication and cybersecurity applications.

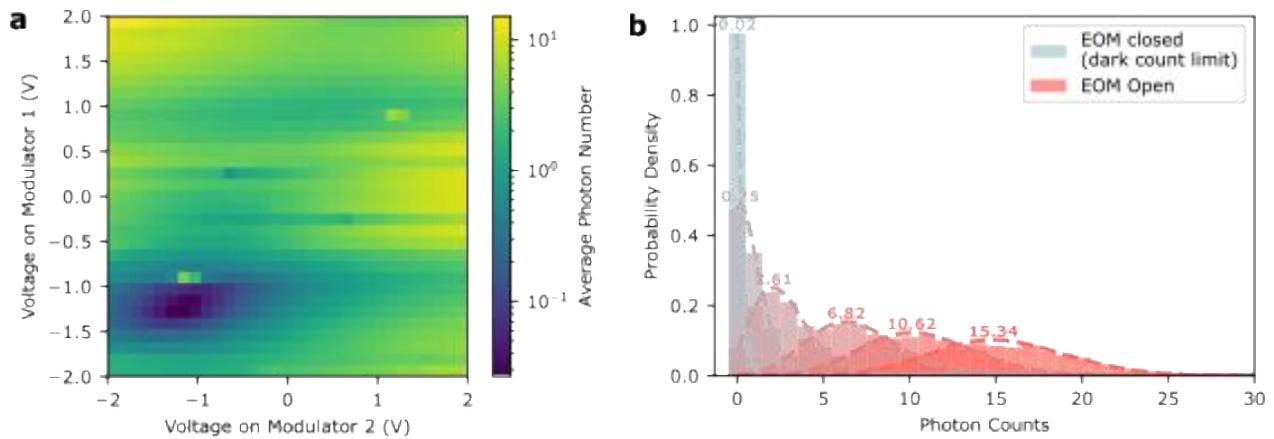

**Figure 5. Single photon Poisson statistics** a) Logarithmic plot of the average output photon number depending on voltage applied to each of the two modulators. The counting interval for each voltage setting is 100 μs. b) Probability density of SNSPD counts in 100 μs time interval for different voltage setting of the amplitude modulator. Dashed lines indicate a fitted Poissonian distribution, with the mean indicated at the maximum of each Poisson curve.

The tunability range in our setup is bounded by the SNSPD performance. On the low end, the detector's dark count rate of 200 cps sets a noise floor of approximately 0.02 counts per 100μs. On the high end, latching of the SNSPD needs to be prevented. In order to not overcome the latching limit, the attenuation was set to an average maximum of 15 photons (corresponding to 150 kcps) per interval to keep safe distance from latching behaviors. Within this dynamic range, we achieve precise and reproducible control over the statistical behavior of the light using only the modulator's drive bias.

## Discussion

The ability to modulate the statistical distribution of light in real time introduces a new operational paradigm in integrated photonics that complements conventional control over amplitude, phase, and wavelength. Because the EDFA operates in deep saturation, the mean output power is effectively pump-limited and therefore remains stable during statistical transitions. Suppressing or transmitting the seed primarily alters the ratio of spontaneous to stimulated emission without significantly affecting the average output power[42]. By enabling programmable transitions between coherent and bunched light at GHz speeds, the photon statistics transducer demonstrated here fills a long-standing gap in reconfigurable optical systems. While previous works have shown static or passively tunable spectral shaping of ASE sources, to the best of our knowledge, no prior system has demonstrated voltage-controlled switching between Poissonian and super-Poissonian photon statistics using only integration-compatible components. Our approach uniquely combines electro-optic attenuation, coherent seeding, and gain dynamics to achieve programmable statistical modulation without changing the light source or introducing mechanical elements. This positions the photon statistics transducer as a practical functional primitive for future integrated photonic systems, especially in hybrid classical-quantum architectures where statistical control is essential.

The continuously tunable regime between $g^{(2)}(0) = 1$ and $g^{(2)}(0) = 2$ is of particular importance. In this range, the optical field exhibits a controlled mixture of coherent and ASE components, allowing the entropy and correlation structure of the light to be precisely set by an electrical drive. This capability is distinct from simple on/off switching, as many applications require intermediate statistics: stochastic photonic processors depend on adjustable noise levels[2,9], secure optical systems benefit from tunable statistical masking[43], and classical–quantum hybrid architectures exploit graded photon-number fluctuations for adaptive computation[24]. By enabling deterministic, voltage-programmable access to these intermediate states, the demonstrated architecture provides a functionality not achievable with conventional source-engineered emitters.

The significance of this capability extends beyond the specific device architecture. Our results show that photon statistics can be modulated deterministically using only electronic drive signals, with no change to the optical source or receiver. The observed second-order coherence $g^{(2)}(0)$ is tunable from 1.0 to 1.7, offering a powerful degree of freedom for applications that require precise entropy management or stochastic behavior. The specific range $1 \leq g^{(2)}(0) \leq 1.7$ observed here is set by the multimode nature of ASE and by the finite optical and electrical detection bandwidth. The demonstrated architecture itself is not intrinsically limited to this range;

integration with single-mode erbium-doped waveguide amplifiers or non-classical emitters would enable access to thermal statistics with $g^{(2)}(0) \approx 2$ as well as sub-Poissonian regimes[42]. It has to be noted, that the maximum obtainable $g^{(2)}(0)$ depends on the mode degeneracy $M$ of the EDFA[8,44]. Moreover, the transition occurs on a nanosecond timescale, established by the electrical bandwidth of the TFLN modulator, which makes the transducer compatible with high speed signaling formats, adaptive photonic logic, and time resolved photon level sensing.

Functionally, the photon statistics transducer can be viewed as an optical entropy engine: it enables the local injection, suppression, or shaping of noise in integrated systems, all in response to an electrical signal. This makes it a valuable building block for quantum key distribution (QKD), physical random number generation, and statistical masking in secure communication links. In neuromorphic and reservoir computing, where controlled randomness plays a critical role in training dynamics or emulating stochastic neural behavior, the ability to modulate the photon-number distribution directly within the optical domain introduces a new layer of system programmability. In all of these cases, the availability of entropy is not a side effect but a required resource[9,17], and the transducer offers it on demand.

Prior approaches to tunable photon statistics have largely relied on resonant, cavity-based, or gain-dynamic mechanisms. Fano-resonance interference in photonic waveguides allows static reshaping of photon-number distributions but is spectrally narrow and not electrically tunable[32]. Chaotic semiconductor lasers and instability-driven systems can generate bunched or partially coherent light, yet the statistical properties depend on internal dynamics and lack deterministic voltage control[33]. Atom-based interferometers and cavity-QED platforms can tailor photon statistics with high precision but require complex setups and are not directly compatible with integrated photonics[4,5]. In contrast, the method demonstrated here leverages electro-optic seed suppression in a saturated EDFA to achieve nanosecond-scale programmability of photon-number distributions. This mechanism is broadband, requires no resonant enhancement, and uses only standard integrated components (InP laser, TFLN modulator, fiber amplifier), making it compatible with wafer-scale photonic integration and scalable system design.

Importantly, the architecture is designed from the ground up with scalability in mind[45]. All components used in this work are either commercially available in integrated form or have already been demonstrated in chip-scale platforms, including the InP laser[46], the TFLN modulator[26,47], on-chip Erbium amplifiers[48], and superconducting single-photon detectors[49]. As such, the photon statistics transducer is not merely a laboratory demonstration, but a practical candidate for inclusion in reconfigurable photonic circuits, especially in applications that span both classical and quantum domains.

Bias stability is an important consideration for long-term operation of interferometric electro-optic modulators. In the present device, the relevant working point corresponds to deep extinction of the coherent seed before amplification, rather than operation at a precise quadrature phase. Once the seed is suppressed by more than 40–50 dB, small thermal or environmental drifts of the modulator phase have little influence on the resulting photon statistics, since the ASE contribution dominates the output. In addition, modulators can be designed such that the maximum suppression state corresponds to the zero-voltage ("ground state") condition, which significantly reduces sensitivity to bias drift during operation. This can be achieved, for example, by incorporating thermal tuning elements that set the interferometer phase bias during calibration. For extended operation or field deployment, several bias drift mitigation strategies have been demonstrated. These include fabrication-level approaches[50,51] as well as active stabilization methods such as slow feedback loops based on monitoring a weak optical tap[52]. Such techniques are routinely used in thin-film lithium niobate modulators and are fully compatible with the cascaded-MZI architecture used here.

While the present demonstration focused on transitions from coherent to thermal statistics, driven by the interplay between laser seeding and ASE in an EDFA, the same core principle can be extended. The modulator's ability to suppress light down to the sub-photon level already enables high-resolution shaping of photon-number distributions. With the integration of a true single-photon source[53], the same architecture could support transitions into sub-Poissonian regimes as well, enabling future exploration of squeezed or anti-bunched light modulation on-chip. This would establish the transducer not just as a source controller, but as a statistical interface between quantum and classical photonic layers.

It is worth noting that the spectral compatibility of the system is currently limited by the operating bands of integrated EDFAs and InP lasers, which primarily cover the O-, C-, and L-bands. Nevertheless, the underlying principle of statistical modulation is agnostic to wavelength, and with appropriate amplifier designs or source materials, the transducer concept can be extended to visible and near-infrared regimes that are of particular interest for atomic systems and visible-wavelength quantum memories.

The concept demonstrated here is not restricted to fiber-based gain media. The same statistical-control mechanism can be implemented using integrated waveguide amplifiers[54,55], including semiconductor optical amplifiers and rare-earth–doped gain sections, which can be co-fabricated[56] or heterogeneously bonded[57] with thin-film lithium niobate modulators. Such on-chip integration would enable a compact, fully programmable source of tunable photon statistics. Moreover, the architecture is compatible with quantum or nonclassical input fields. When seeded by sub-Poissonian sources, a high-extinction electro-optic control stage can preserve or reshape $g^{(2)}(0) < 1$ statistics, providing an avenue toward electrically programmable control of both classical and quantum photon-number distributions. These integrated and quantum-compatible pathways broaden the scope of the platform and suggest several promising directions for future devices.

Overall, this work introduces a new functional primitive for integrated photonics by enabling electro-optic control over photon-number statistics and establishing a foundation for devices that operate not only on optical signals but on their statistical character, with entropy emerging as a programmable resource. The key advance lies not in the well-established high-speed phase control of a TFLN MZI, but in the system-level capability to electrically program photon statistics through high-extinction seed suppression, which transforms the modulator from a phase-control element into a means of deterministic statistical transduction.

## Methods

**Device Fabrication and Architecture**

The core component of the photon statistics transducer is a high-extinction amplitude modulator fabricated on a TFLN platform using foundry-based processes. The device comprises two integrated Mach-Zehnder interferometers (MZIs) that are interferometrically coupled on the same chip, forming a compound architecture in which the optical output of one modulator influences the interference condition of the other. This topology enables deterministic shaping of the output field through multiple internal phase shifts and delivers extinction performance that surpasses that of a single-stage modulator.

The TFLN chip was fabricated via standard x-cut LNOI processes and is based entirely on components available in the foundry's photonic design kit (PDK). Both MZIs utilize directional couplers drawn from the standard library, which are nominally designed for 50:50 power splitting. However, to compensate for potential fabrication-induced deviations, the interferometric coupling between the two MZIs was introduced deliberately, allowing the overall system to tolerate significant asymmetry in individual coupler performance. This resilience is a key architectural feature: even if the internal splitting ratios differ from the design target, the system retains its ability to achieve deep destructive interference, as described in the Results section.

Each MZI includes push–pull traveling-wave electrodes patterned via lift-off metallization. The electrodes are driven using synchronized RF signals applied to the chip through wire bonds, allowing modulation of both phase arms. The measured half-wave voltage $V_\pi$ is approximately 3.4 V at 1550 nm, as shown in Figure 3. Due to the coherent coupling between the two interferometers, the modulators cannot be driven independently: the setting of one stage influences the output of the full system. This interdependence provides an additional degree of tunability and is exploited to maximize extinction.

The modulator is fabricated as a bare die and mounted on a printed circuit board (PCB) for RF control. Electrical contacts are made via wire bonding, and optical access is achieved through surface grating couplers optimized for the telecom C-band. The chip is optically probed using fiber arrays in non-packaged configuration during measurements. No active thermal tuning or bias stabilization was used. The extinction ratio was determined by direct measurement of the transmitted optical power at the output using a calibrated photodiode and power meter. Under optimized bias conditions, the system achieved a static extinction ratio of 51 dB, as reported in the Results section.

The optical input to the photon-statistics transducer is provided by a distributed-feedback (DFB) laser fabricated on an InP platform and sourced from a commercial photonic foundry. The device operates in the C-band with a typical linewidth below 80MHz, includes an integrated heater for wavelength tuning, and is wire-bonded to a carrier PCB without active cooling. Edge coupling is implemented via spot-size converters to an angle-polished single-mode fiber array, aligned using a six-axis positioning stage (H-206, Physik Instrumente (PI) SE & Co. KG). A source meter (Keithley 2450, Tektronix Inc.) supplies the drive current. As an integration-ready component, the InP laser serves as a representative seed source for future monolithic or hybrid photonic systems. A schematic of the setup, along with photographs of the bare-die TFLN modulator and packaged InP laser, is shown in Figure 1.

**Experimental setup for photon statistics modulation**

To characterize statistical transitions in the high-power regime, the output of the TFLN amplitude modulator was routed through a low-noise EDFA (LNFA-33, PriTel Inc.) with a small-signal gain of 33 dB. The EDFA amplified the modulated light depending on the applied voltage to the transducer: when open, it transmitted and amplified the seed laser signal, producing coherent output; when closed, it emitted only amplified spontaneous emission (ASE), corresponding to thermal-like light with bunched photon statistics. To prevent photodiode saturation or damage in the high-gain regime, a 10 dB inline optical attenuator was inserted immediately after the EDFA output. During all measurements, the EDFA was operated in a strongly saturated regime. As a result, its output power was pump-limited and remained essentially constant irrespective of the seed level. The seed power (<100 µW) was more than 20 dB below the EDFA output (~30 mW), so seed suppression affected only the statistical composition of the output without introducing appreciable changes in average power.

It is important to note that the nanosecond-scale transition between Poissonian and super-Poissonian output is not limited by the intrinsic gain dynamics of the EDFA. The erbium population remains quasi-static on the microsecond scale, and the statistics are instead governed by the instantaneous balance between seeded stimulated emission and unseeded ASE[42,58]. Stimulated emission responds at the timescale of the optical field, permitting changes in output photon statistics at the full modulation bandwidth of the TFLN MZI[59]. In this regime, the EDFA acts as a pump-limited noise source whose statistical composition is determined solely by the seed-suppression ratio, which can be varied at multi-GHz speeds independent of the erbium relaxation time.

Intensity fluctuations of the EDFA output were measured using a 38 GHz AC-coupled photodiode with integrated transimpedance amplifier (RXM38AF, Thorlabs Inc.). The TIA was configured to a gain setting of 900, and the detector output was directly connected to a high-speed oscilloscope (Infinium DSA-X 95004Q with 62 GHz upgrade, Keysight Technologies Inc.). The oscilloscope sampled the analog voltage traces at 80 GSa/s for the autocorrelation and RMS noise measurements, and at a down sampled rate of 256 MSa/s for the RIN analysis. No additional filtering was applied beyond the AC coupling and internal scope bandwidth settings.

The RMS noise amplitude was computed from the time-domain photodiode voltage trace over a window of 4 ms, and plotted as a function of the applied modulation voltage. This allowed direct observation of the transition from low-noise (stimulated) to high-noise (ASE-dominated) output, as described in the Results section. Relative intensity noise (RIN) was computed from the same voltage trace via fast Fourier transform (FFT), yielding power spectral densities in dB/Hz. The difference between the fully open and fully blocked transducer states amounted to 2.38dB/Hz, confirming increased statistical fluctuation when the EDFA operated without a seed.

To probe the photon statistics indirectly, we computed the second-order autocorrelation $g^{(2)}(0)$ from the photodiode voltage using the normalized second moment. In the coherent regime, the measured value was, while in the noise-dominated regime it increased to $g^{(2)}(0) = 1.661 \pm 0.098$. Although the ideal thermal limit $g^{(2)}(0) = 2$ was not reached due to finite optical and electrical bandwidth constraints, the observed transition confirms the modulation of photon statistics via the EDFA path. The photodetector bandwidth was limited to 38 GHz, while the ASE source bandwidth was constrained to approximately 10 GHz, resulting in partial temporal filtering of the intensity fluctuations. Nonetheless, the combined RMS, RIN, and $g^{(2)}(0)$ data clearly demonstrate the statistical transition from Poissonian to super-Poissonian regimes.

**Experimental setup for single-photon number modulation**

To resolve the photon-number statistics of the attenuated laser output, a single-photon detection arm was implemented using a fiber-coupled, waveguide-integrated SNSPD. The optical signal, modulated by the dual-stage TFLN amplitude modulator, was further attenuated by a fixed 17 dB optical attenuator before entering the SNSPD input. This allowed the mean photon number per time interval to be tuned between the sub-photon regime and the detector's maximum sustainable count rate, with all attenuation performed in-line to preserve timing integrity.

The detector used was a DENA Desktop system by Pixel Photonics, which integrates SNSPDs and cryogenic electronics into a compact commercial package. The system detection efficiency at 1550 nm was specified at 29%, and the system included an internal low-noise amplifier stage. The output was connected to a time-tagging module (Time Tagger Ultra, Swabian Instruments GmbH), which recorded timestamped photon detection events with a timing resolution of 8 ps. The SNSPD was operated at the systems base temperature of 2.4 K.

To construct photon count histograms, time intervals of 100 μs duration were defined as virtual time gates using the software control of the time tagger. A dedicated virtual channel was triggered by the applied drive voltage of the TFLN modulator, allowing conditional counting only when the modulator was biased at a defined state. This approach enabled photon number statistics to be sampled synchronously with the voltage-controlled state of the transducer. Photon counts were accumulated over many intervals to extract the mean photon number ⟨$n$⟩, as well as the full photon-number distribution.

Data acquisition and post-processing were performed using a custom Python-based software stack. The time-tag data was processed into photon count histograms per modulation state, and each distribution was fitted to a Poisson model to confirm coherence at low power. As described in the Results section, the observed distributions transitioned from sparse Poissonian behavior to broader Gaussian-like shapes as the input power increased, demonstrating the transducer's continuous tunability across the photon statistics landscape.

# Data Availability

The authors declare that all data supporting the findings of this study can be found within the paper and its Supplementary information files. Additional data supporting the findings of this study are available from the corresponding author upon reasonable request.

# Funding

This research was supported by the European Union's Horizon 2020 research and innovation programme (grant no. 101017237, PHOENICS project) and the European Union's Innovation Council Pathfinder programme (grant no. 101046878, HYBRAIN project). We acknowledge funding support by the Deutsche Forschungsgemeinschaft (DFG, German Research Foundation) under Germany's Excellence Strategy EXC 2181/1 – 390900948 (the Heidelberg STRUCTURES Excellence Cluster), the Excellence Cluster 3D Matter Made to Order (EXC-2082/1—390761711) and CRC 1459 'Intelligent Matter'.

# Acknowledgements

We thank Jochen Stuhrmann, from Illustrato for assistance with the illustrations.

# Author contributions statement

Concept: J.R.B, F.B.P. Methodology: J.R.B., Liam M., A.V. Investigation: J.R.B., R.P., J.R., F.E., Lennart M., J.B. Visualization: J.R.B., X.M., Funding acquisition: W.H.P.P., X.M. Project administration: W.H., W.H.P.P., X.M. Supervision: F.B.P., W.H., W.H.P.P., X.M, Writing—original draft: J.R.B., X.M. Writing—review and editing: All authors.

# Ethics Declaration

## Conflict of interest

W.H. and W.H.P.P. are involved in developing single photon detectors at Pixel Photonics GmbH. J.R.B., J.R. and W.H.P.P. are involved in developing electro-optic modulators at Linq Photonics GmbH. The remaining authors declare no competing interests.